\pdfoutput=1
\RequirePackage{fix-cm}
\documentclass[smallextended]{svjour3}       % onecolumn (second format)
\smartqed  % flush right qed marks, e.g. at end of proof
\usepackage{graphicx}
\begin{document}

\title{Machine learning as an instrument for data unfolding}

\journalname{DESY-17-214}


\author{Alexander Glazov      
}


\institute{Alexander Glazov \at
              DESY, Hamburg, Notkestraße 85 D-22607 Hamburg \\
%              Tel.: +123-45-678910\\
%              Fax: +123-45-678910\\
              \email{alexander.glazov@desy.de}           %  \\
}

\date{ }
%Received: date / Accepted: date}

\maketitle

\begin{abstract}
  A method for correcting for detector smearing effects using machine learning techniques is presented.
  Compared to the standard approaches the method can use more than one reconstructed variable
  to infere the value of the unsmeared quantity on event by event basis. The method is implemented
  using a sequential neural network with a categorical cross entropy as the loss function. It is tested on a
  toy example and is shown to satisfy basic closure tests. Possible application of the method for analysis of the
  data from high energy physics experiments is discussed.
\end{abstract}

\section{Introduction}
\label{intro}
The problem of unfolding, i.e. correcting the observed distributions in data for the detector resolution effects in order to obtain the true underlying distribution of the physical quantity, is a common problem in high energy physics (HEP). A number of methods has been developed to solve it
which typically formulate 
the problem in terms of the transfer matrix from the truth to reconstructed distribution. The unfolding then corresponds to an inversion of the transfer matrix. A simple inversion, however, may lead to a significant bin-to-bin anticorrelations for the unfolded distribution, spoiling stability of
the result~\cite{Anykeyev1991,Blobel:2002pu}.
A number of regularisation methods has been proposed to overcome this problem.
They use smoothness of the physical observable as an extra prior information~\cite{Zhigunov:1983ee,Blobel:2002pu,tikhonov}, introduce explicit priors using e.g. reduced cross-entropy method~\cite{Schmelling:1993cd}, analyse the transfer matrix with singular value decomposition procedure and keep significant eigenvalues only~\cite{Hocker:1995kb}, invert the transfer matrix in an iterative approach and truncating the number of iterations~\cite{Shepp82,Kondor83,Multhei87,DAgostini:1994fjx,Malaescu:2009dm}. 
There are several other approaches to solve the unfolding problem, e.g. by applying extra smearing
corrections to the reconstructed data, obtaining this way spectra which are folded multiple times with the detector response, and extrapolating them back to zero foldings~\cite{Monk:2011pg}. An application of machine learning methods for unfolding is discussed in Ref.~\cite{Gagunashvili:2010zw}.
A recent review and performance comparison of different unfolding methods used frequently
in HEP can be found in Ref.~\cite{Schmitt:2016orm}.

In this letter, the problem of unfolding is viewed as a categorization problem,
i.e. assigning for each reconstructed data event the most probable truth bin.
Similarly to the approach described in Ref.~\cite{Lindemann:1995ut}, the method does not require
binning for the reconstructed quantities, the correction is applied on event-by-event basis,
and the unfolding is performed in an iterative procedure.
A set of closure tests is proposed to check convergence of the procedure and that it is bias free.

The categorization is a common problem for the modern machine learning (ML) methods. 
For example, the standard benchmark for the ML algorithms, the classification
of the hand-written digits using MNIST database~\cite{mnist},
is a categorization problem. A number of methods has
been developed to solve it with high efficiency which can be employed for the unfolding problem.
These include methods such as boosted decision trees and artificial neural networks.
Deep neural networks  (DNN) are under active development now in part due to advances in computational resources which are
suitable for them.
DNNs with various architectures  can be tried for the unfolding problem. In particular, networks with
convolutional layers may explore distance relation between variables and truth-bins. Sequential
networks with fully connected layers can be used for generic unfolding problems which include long-range migration effects due to e.g. kinematic misreconstruction.

The strength of the ML methods is that they can use large amount of information as input.
The ultimate goal of ML  in HEP is to determine the underlying physics quantities directly from the unprocessed detector information. Given the current status of algorithms and computing resources
the goal is still to be reached in future. 
It is also important to see first how the method performs in comparisons to other approaches using a single input variable. The method is introduced in the following using a  one-dimensional unfolding problem, starting from no smearing and miscalibration effects, to check that the model is sufficient.
The miscalibration and smearing are added next leading to an iterative unfolding procedure.  The second reconstructed variable is introduced as the final example.
The letter concludes with a discussion of the results and possible future applications of the method.

\section{Method description}
The described below example and the unfolding procedure,
called in the following as ``ML unfolding'', can be found on {\tt github}\footnote{https://github.com/aglazov/MLUnfold.git}
as a python jupyter notebook. The code uses Keras~\cite{chollet2015keras}  and borrows the underlying
neural network architecture from the  example files of the package.

The classifier used for the unfolding is based on a three layer sequential neural network.
The first layer
contains $N_{\rm inp}$ neurons where $N_{\rm imp}$ is the number of input variables.
For a standard unfolding problem $N_{\rm inp}=1$, but $N_{\rm inp}=2$ is examined in this letter as well.
The second layer contains 
$\kappa N_{\rm bin}^2$ fully connected dense cells employing ReLU activation function,
$f(x) = \max(0,x)$. Here $N_{\rm bin}$ is the number of truth bins and $\kappa$
is determined empirically to be $\kappa \sim 8$.
The second layer plays the role analogous to the transfer matrix, which justifies a large number of neurons, proportional to $N_{\rm bin}^2$.
The ReLU activation function is well suited for the first task of the neural network to determine the bin boundaries for the reconstructed variable. The final third layer contains $N_{\rm bin}$ categorical neurons activated using the softmax function, $ f_i(\vec{x}) = \exp x_i / \sum_{k=1}^{N_{\rm bin}} \exp x_k$. The network uses categorical
cross entropy, $H(\vec{p},\vec{q}) = -\sum_{e=1}^{N_{\rm evt}} \sum_{i=1}^{N_{\rm bin}} p^e_i \log q^e_i$  as the loss function, where $p^e_i$ and $q^e_i$ are the truth and predicted
probabilities for an event $e$ and summation runs over all $N_{\rm evt}$ Monte Carlo events and truth bins.
By construction, $p^e_i =1$ for one $i$th-bin and zero for all other bins while $q^e_i$ may be non-zero for a couple of bins, as determined by the softmax activation function. The training of the network employs the ADADELTA method
which uses  dynamically optimized per-dimension learning rate for the gradient decent
minimization~\cite{Zeiler2012ADADELTAAA}.

It is essential to ensure that the unfolding procedure is not biased towards the distribution used in the simulation to train the neural network.
Therefore the ML unfolding is performed iteratively. The classifier is trained initially using an input distribution of $x=x_g$ which may have
no prior information, so called flat prior $F(x_g) = const$, or be based on previous measurements and theoretical expectations.
The trained classifier is then applied to the data, to determine $q^{e,{\rm data}}_i$ that is used to compute an
updated binned $F'_i(x)$ distribution as $ F'_i(x) = 1/N_{\rm evt,\,data} \sum_{e=1}^{N_{\rm evt,\,data}}q^{e,{\rm data}}_i $,
which in turn is employed to re-sample the Monte Carlo simulation and re-train the
classifier. The procedure is repeated until convergence which can be detected based on stability of the unfolded result over several consequent iterations compared to statistical
uncertainties.
Given that the model used for the ML unfolding has many redundant parameters, many possible
training results can be obtained at each iteration.
This introduces additional statistical uncertainty which can be removed by taking an average over
unfolded results using several  ML unfolding sessions that use different input random numbers. Twenty sessions are used for the analysis
presented here.

\begin{figure}[tbh]
\centerline{
\includegraphics[width=0.6\linewidth]{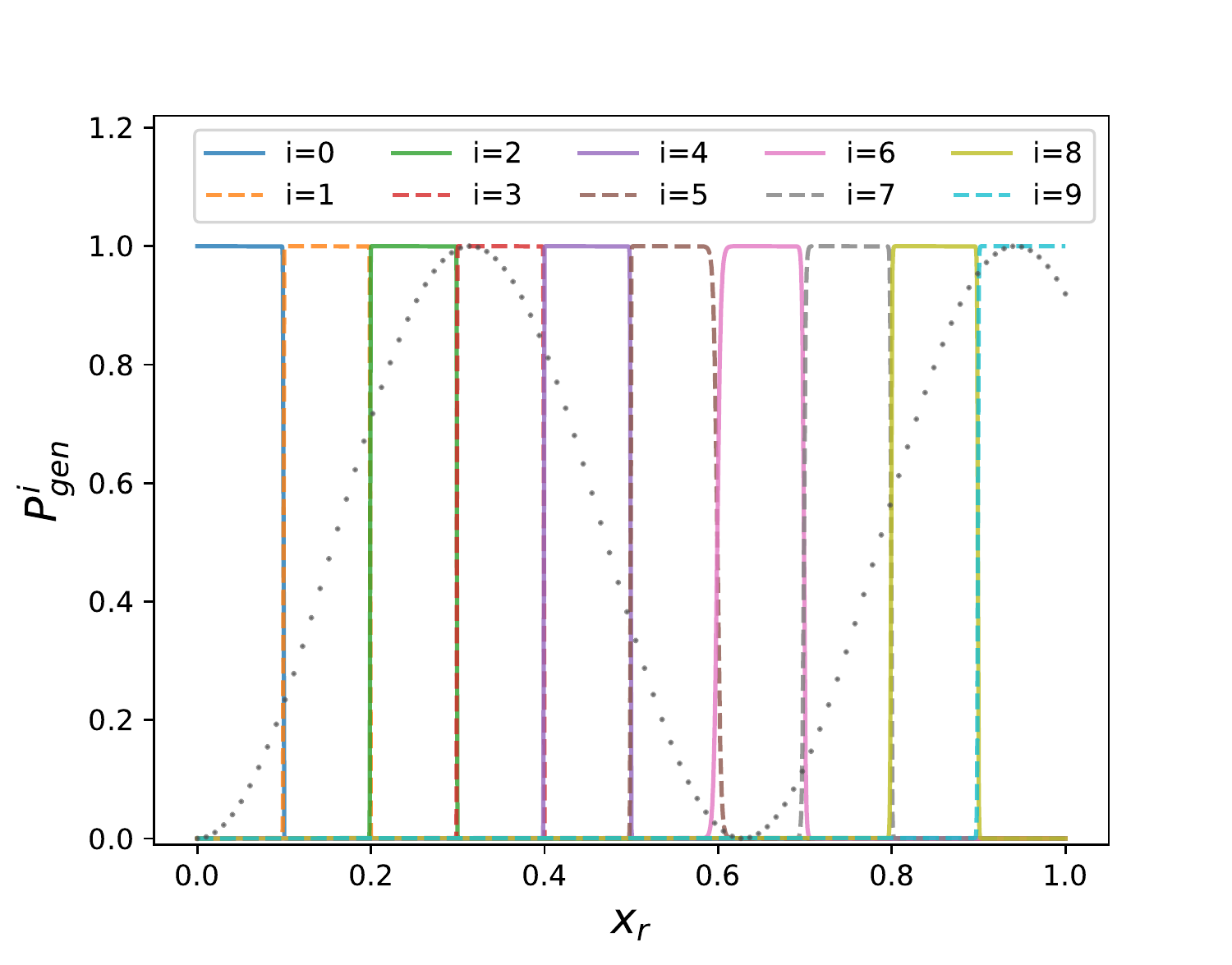}}
\caption{\label{fig:binboundary}Probability for the consecutive truth bin number, $P^i_{gen}$, shown with
different colours and line styles,
as a function of the reconstructed variable $x_r$ for the simulation without smearing and ML estimator trained based on the truth prior, which is shown as a dotted line.}
\end{figure}
The described above iterative procedure is a common method for solving unfolding problems and has been proved to converge for a wide class of
them~\cite{Shepp82,Kondor83,Multhei87}. The convergence can be probed for the ML unfolding using a set of closure tests.
The zero-level test verifies that the classifier internal architecture is sufficient to solve the classification problem
and determine the truth-bin boundaries in the trivial case when reconstructed $x_r = x_g$. The test should be fulfilled with any
prior distribution  after the zeroth iteration. The test establishes the required ML topology and the training strategy, such as the number
of batches and epochs used for the training. When performed using the truth-based prior,
it also checks the sufficiency of the Monte Carlo sample statistics, especially for the bins correponding to the minimum of the distribution.

The first-level  test checks that the setup provides a fixed-point for the iterative procedure when using the
truth-based prior.
Agreement between the input truth and the unfolded distribution is expected
after the zeroth iteration which should hold within statistical uncertainties.

The final closure test is based on different prior
distributions for the training and validation samples and is used to determine required number of iterations.
For example, a flat prior can be used for the initial training.
A convergence should be observed for a range of validation distributions.
This test can be performed in a more  systematic manner if 
the expected distribution has well-defined uncertainties. In this case, the prior distribution can be based on the expected distribution and
varied  within its uncertainties
to verify convergence to the central value, which should be achieved to a much better level than the size of the uncertainty. 
It could be also instructive to verify that the truth-based prior is a stable fixed point by altering the prior slightly and verifying
that the procedure converges rapidly.

As an example, the ML unfolding is tested using a distribution following $F(x_g) = \sin^2 5x_g$ function for the generated
$x_g$ varied
between $0$ and $1$.  $N_{\rm bin} = 10$ truth bins are used for the unfolding which are distributed uniformly
with a bin width of $0.1$. The efficiency of the detector is assumed to be uniform in $x_g$,
focusing the test on the bin migration effects. 
Training and test samples with $2 \times 10^{6}$ and $20 \times 10^3$ random events are used, respectively.
The test sample is use as pseudo-data for  unfolding.
Two initial training samples are generated following the truth-based $F(x_g)= \sin^2 5x_g$ and flat $F(x_g) = const$ priors.

The zeroth-level check of the procedure is to ensure that the neural network is sufficiently large to determine truth bin boundaries for the case without detector smearing, i.e. when reconstructed $x_r = x_g$. Indeed, in this case the neural network reaches $>99.9\%$ prediction accuracy $P_A$, determined by comparing the reconstructed bin with maximal probability to the truth bin, 
for the validation sample after training. For the prior with non-uniform distribution, significant amount of the training cycles is required to achieve this accuracy. The result of the training using batch size of $1000$ events and $400$ epochs
is shown in Figure~\ref{fig:binboundary}. Very sharp bin boundaries are visible for all but the bin corresponding to the minimum of the prior distribution, but even for this bin the accuracy is sufficient, considering typical smearing effects.
\begin{figure}
\hspace*{-1cm}\includegraphics[width=1.1\linewidth]{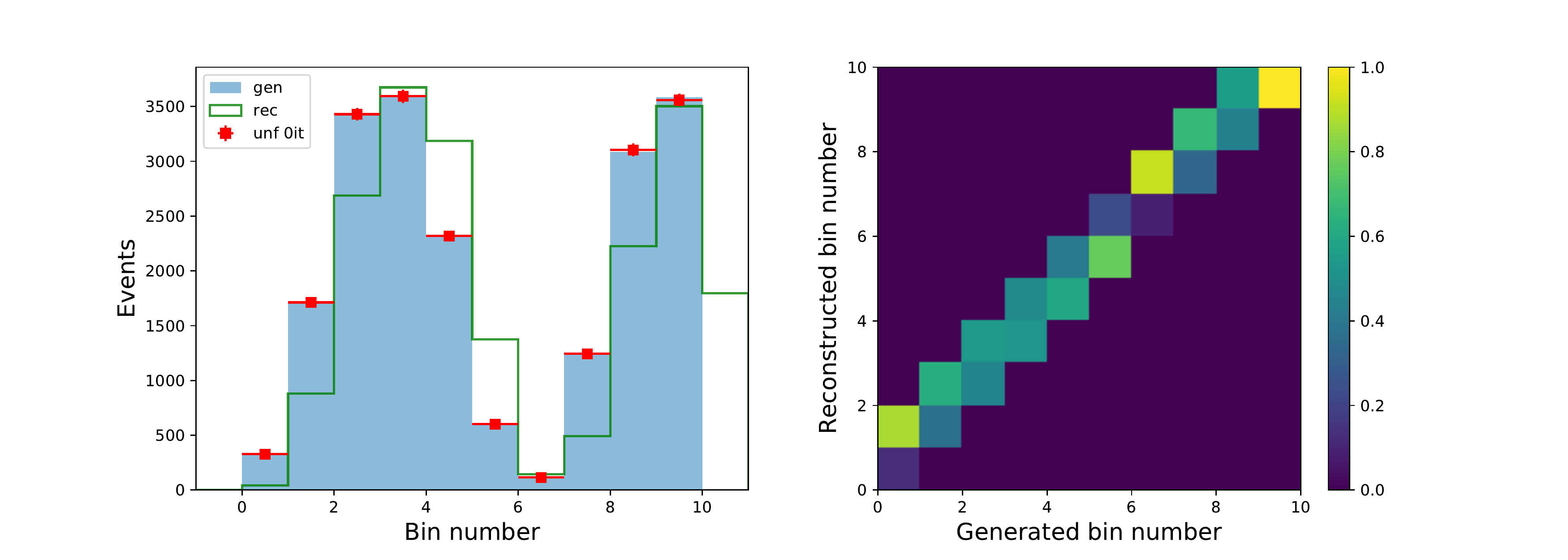}
\caption{\label{fig:figshift} Left: comparison of the truth, reconstructed and unfolded distributions for
the test when the reconstructed $x_r$ is shifted vs truth $x_g$ as $x_r = x_g + 0.05$. Right:
corresponding response matrix with the over-flow events in reconstructed bin number  placed in the
last  bin. }
\end{figure}

This is a non-trivial result already since the network did not receive directly any information on the bin boundaries. It can be extended to the first interesting application: consider bias of the reconstructed variable $x_r = x_g + 0.05$. It is illustrated in Figure~\ref{fig:figshift}. A sizable shift of the reconstructed distribution with
respect to the truth is
observed in this case, making simple bin-by-bin unfolding inapplicable. The shift introduces large
off-diagonal elements to the transfer matrix.
The ML unfolding, trained using the flat prior with no
iterations performs very well, reproducing the truth within statistical uncertainties of the test sample which 
estimated using bootstrap resampling method~\cite{efron1979}. The bootstrap method uses generated events
multiple (or zero) times following Poisson probability distribution with the expectation value $\mu=1$. Thirty
bootstrap replica are used in the analysis presented here.

\begin{figure}
\hspace*{-1cm}\includegraphics[width=1.1\linewidth]{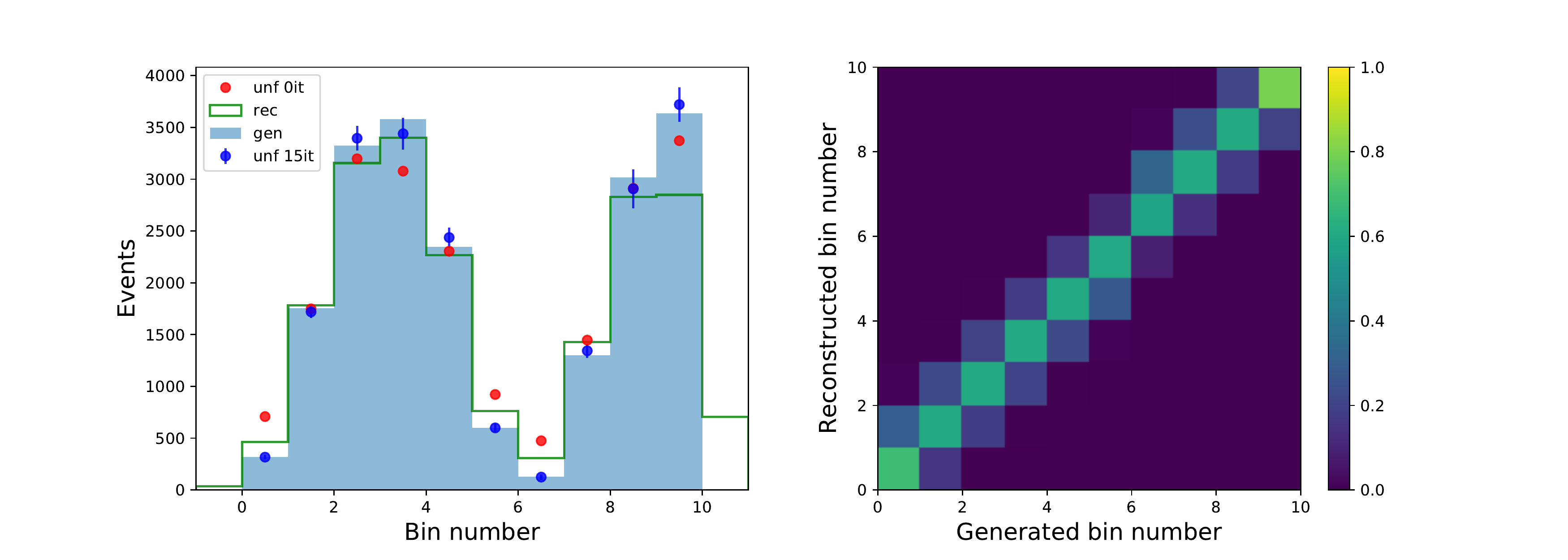}
\caption{\label{fig:figsmear} Left: comparison of the truth, reconstructed and unfolded distributions with
zero and fifteen iterations from the flat prior for
the test when the reconstructed $x_r$ is smeared by $0.05$. Right:
corresponding response matrix with the under- and over-flow events in reconstructed bin number  placed in the first and
last  bin, respectively. }
\end{figure}

\begin{figure}
\hspace*{-0.5cm}\includegraphics[width=1.1\linewidth]{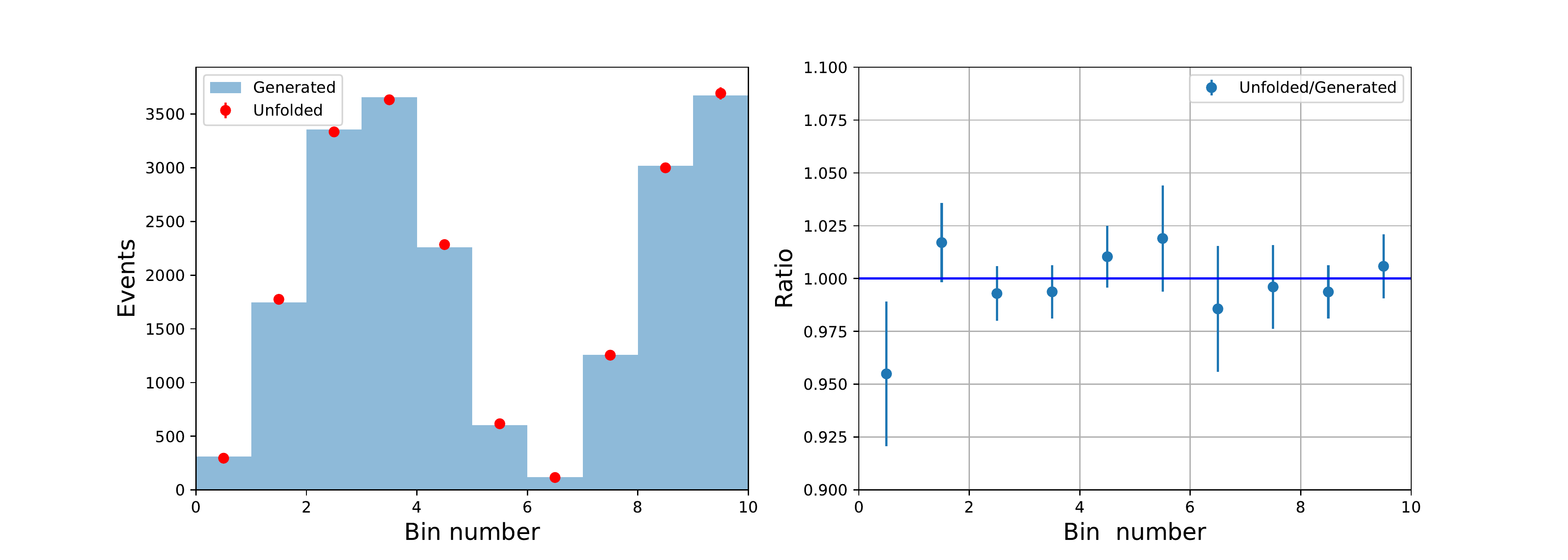}
\caption{\label{fig:xx}Left: comparison of the truth and unfolded distributions after zeroth iteration of unfolding with the prior based
on the truth distribution. Right: ratio
of the truth and unfolded distributions. }
\end{figure}

For the next checks, the bias of the reconstructed variable is removed and half-bin Gaussian smearing ($\sigma = 0.05$) is introduced. This leads to sizable bin-to-bin migrations, see Figure~\ref{fig:figsmear}.
The diagonal elements of the transfer matrix, which show the fraction of events  reconstructed in the same bin as the truth,
the so called bin stability, is at $60\%$ level. The bin purity, which is measured as a fraction of truth events
for a reconstructed bin, is also close to $60\%$, apart from the seventh bin, corresponding to the distribution minimum, where it dips to $20\%$ only. The accuracy of the ML prediction is also at $P_A \sim 64-66\%$ level in this case
with the lower value obtained the flat prior and higher for the truth-based prior. 
It is then verified that when using the truth-based prior, $F(x_g) =\sin^2 5 x_g$, the unfolded distribution agrees with the truth
to within statistical uncertainties of the test sample, see Figure~\ref{fig:xx}.

The flat initial prior is used to train the ML classifier for the following. 
The zeroth iteration of ML unfolding is in fact extra smearing on top of detector resolution effects: with no prior information
on the underlying distribution shape, the neural network can only smear the reconstructed event according to the probabilities to be originated from the truth bins. This can be observed in Figure~\ref{fig:figsmear} by comparing the zeroth iteration to reconstructed distribution.
For the following iterations, the shape of the distribution
obtained from the last iteration is parameterized using cubic splines based on Ref.~\cite{SciPy} and used as a prior for the event generation.
The splines are chosen such that the integrals over bin boundaries agree with $F'_i(x)$ for all bins.

\begin{figure}
\hspace*{-1cm}\includegraphics[width=1.1\linewidth]{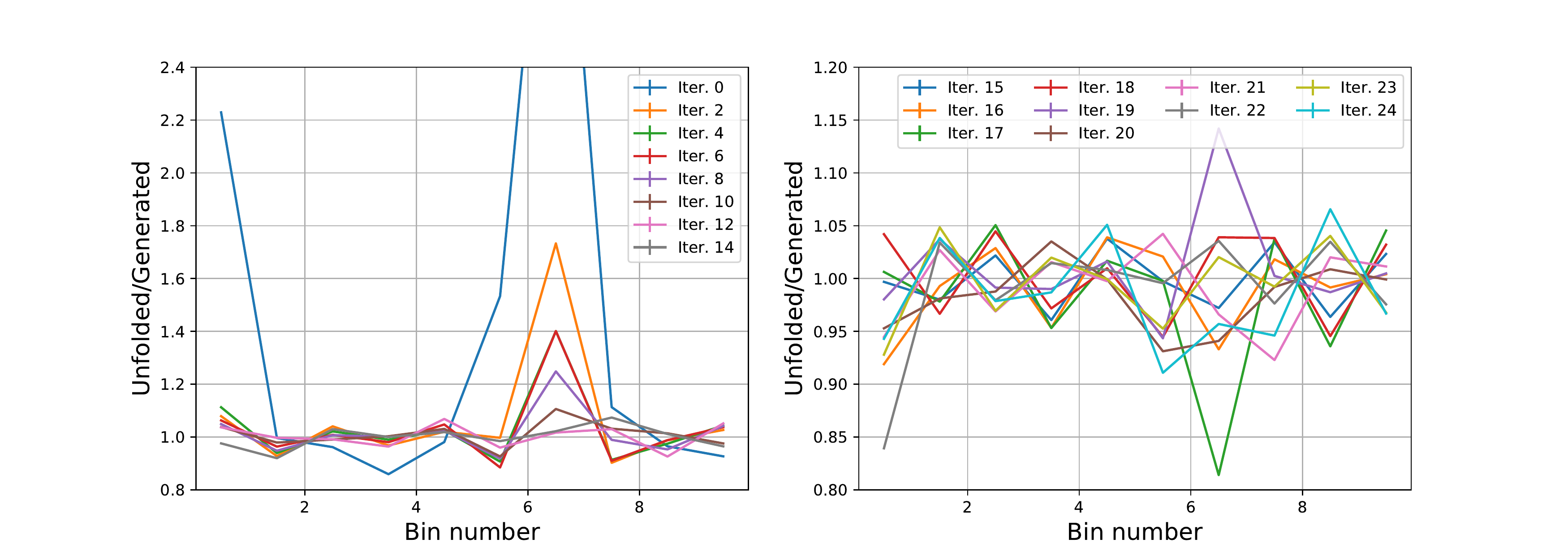}
\caption{\label{fig:converge}
For unfolding based on the single reconstructed variable startng with the flat prior, ratio of unfolded to truth binned distributions for iterations from $0$ to $14$ (left) and from $15$ to $24$ (right). }
\end{figure}

The iterative ML unfolding procedure leads to a fair representation of the underlying truth distribution after the tenth iteration.
The convergence is illustrated in Figure~\ref{fig:converge} which shows rapid convergence up to the tenth iteration and oscillatory behavior for
the following iterations, generating sizable bin-to-bin anticorrelations.

\begin{figure}
\hspace*{-1.cm}\includegraphics[width=1.1\linewidth]{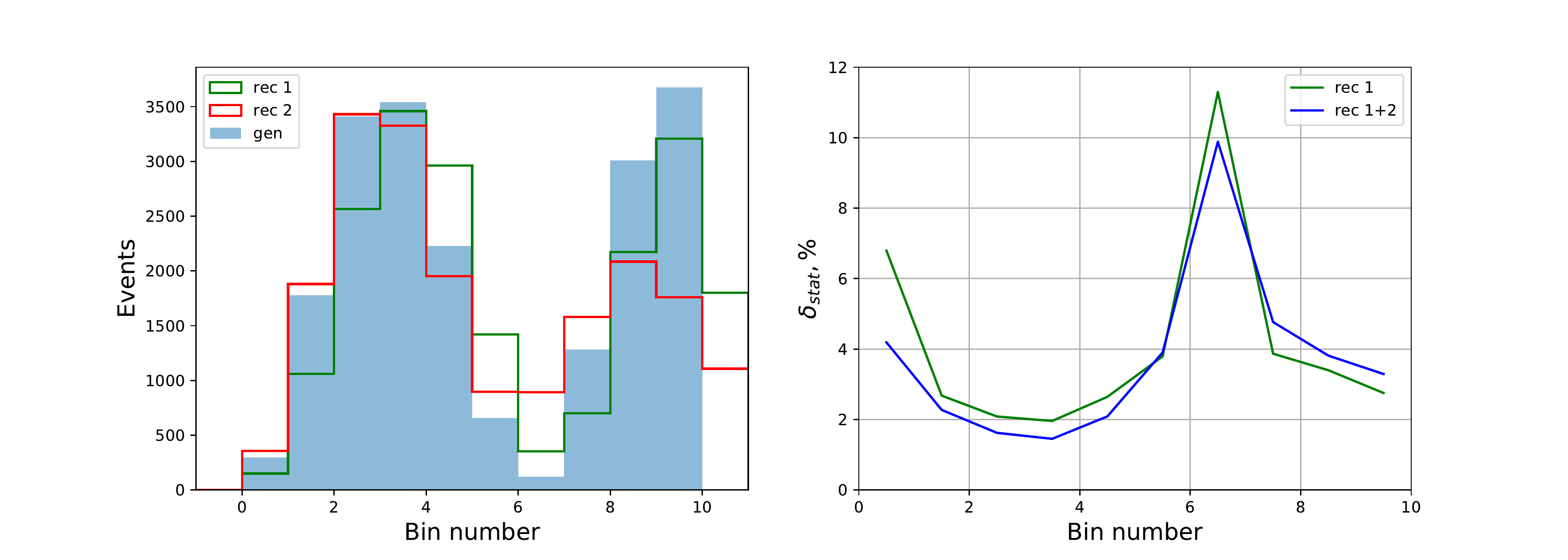}
\caption{\label{fig:third}Left: comparsion of the truth and reconstructed distribution with Gaussian and log-normal smearing labeled as
rec~1 and rec~2, respectively. Right: relative statistical uncertainty of the unfolded result using single Gaussian smeared variable rec~1
and using both rec~1 and rec~2 variables.}
\end{figure}
To illustrate additional capabilities  of the ML unfolding a second reconstructed variable is introduced. The first variable retains Gaussian smearing with $\sigma =0.05$ and additional shift of $+0.05$. The second variable is smeared using log-normal distribution, with $\sigma = 0.15$ and shift of $-0.05$. Difference of the smearing models prevents direct combination of the two variables into a single one.
Figure~\ref{fig:third}, left shows comparison of the generated and reconstructed distributions. The second reconstructed variable has an improved accuracy for the first bins of the distribution, but much worse for the rest.
When trained using both variables, the prediction accuracy of the neural network improves from $P_A \sim 65\%$ to $P_A \sim 70\%$ compared to usage of the
single variable.
Consequently, the  statistical uncertainty  is improved for the two-variable ML unfolding, see Figure~\ref{fig:third}, right.

\section{Discussion}
\label{sec:summary}
A method of machine learning unfolding shows promising performance for the toy example presented here. For a single reconstructed variable, it is similar to iterative approaches with an additional advantage that it does not require binning in the reconstructed quantities. It is insensitive to miscalibration between truth and reconstructed variables and chooses the effective binning automatically. While being much more computationally demanding compared to the standard techniques, the method benefits from many tools developed recently for the machine learning applications,
including hardware optimization. For instance, the training of one epoch for the example discussed in the letter takes $\sim 5$ second of physical time on a Dell Precision 5520 laptop
with Intel i7-7820HQ CPU and  Nvidia Quadro M1200 GPU, thanks to GPU-based acceleration of Keras using the Tensorflow backend. However multiplying this by the number of epochs to complete the training, additional training sessions and bootstrap replica to estimate statistical uncertainties, complete training demands large resources. 

The method can be naturally extended to several input variables. It has been demonstrated that even a variable with a poor resolution can add valuable information for determining the truth distribution. As such, the method can be used as an agregator of the experimental
information to a new variable which can be employed as a proxy to the truth and any other unfolding approach can be applied for the transfer
from this variable to the truth.

The ability of the method to make use of several variables can be beneficial for many types of  measurements. For example,
measurements of the structure functions in deep-inelastic lepton-proton scattering rely on two variables, Bjorken-$x$ and absolute
momentum transfer squared, $Q^2$. These two variables can be reconstructed using the scattered lepton kinematics alone, but they can be also determined based on the hadronic final state (HFS). A number of kinematic methods which combine lepton and HFS information were introduced in the past (see e.g.~\cite{Bassler:1994uq}), and they can be employed by the ML method in an optimal way. Another example is the measurement of the $Z$-boson transverse momentum, $p_{\rm T}^Z$, at the LHC. Currently, two distinct measurements are performed: of $p_{\rm T}^Z$ itself,
which is preferred by theory, and of the variable $\phi^*$ which is related to $p_{\rm T}^Z$ and, being insensitive to the
measured absolute momenta, has higher accuracy experimentally (see e.g.~\cite{Aad:2015auj}). With the ML method discussed here, both
variables can be used to unfold $p_{\rm T}^Z$. One can also use invariant mass of the lepton pair, as the third variable, with a
known $Z$-boson lineshape as a prior, to control calibration and QED radiative effects at event by event level.

To conclude, a ML unfolding method is introduced and tested in this letter. It can serve as a useful tool for several analyses in HEP.

\begin{acknowledgements}
  The author would like to express gratitude to Mateusz Dyndal, Mikhail Karnevskiy,
  Carsten Niebuhr,
  Stefan Schmitt, and Simone Wehle
  for inspiring discussions and comments to the draft. 
\end{acknowledgements}

\bibliographystyle{spphys}       % APS-like style for physics
\bibliography{bibl}   % name your BibTeX data base

\end{document}